**Mobile Phone Use as Sequential Processes:**

**From Discrete Behaviors to Sessions of Behaviors and Trajectories of Sessions**

Tai-Quan Peng[1*], Jonathan J. H. Zhu[2,3]

1. Department of Communication, Michigan State University, U.S.A.
2. Department of Media and Communication, City University of Hong Kong
3. School of Data Science, City University of Hong Kong

[*] Corresponding Author; Email: winsonpeng@gmail.com, Tel.: (+1) 517-355-0221

This article has been accepted for publication in *Journal of Computer-Mediated Communication* published by Oxford University Press.

Acknowledgement:
The study was supported in part by research grants (9360120 and 7004732) from City University of Hong Kong and start-up grant (GE100028) from Michigan State University, respectively.

Authors Biography:
Tai-Quan Peng is an Associate Professor at the Department of Communication, Michigan State University. His current research focuses on computational social science, health communication, political communication, and mobile analytics. (ORCID: https://orcid.org/0000-0002-2588-7491)

Jonathan J. H. Zhu is a Chair Professor of Computational Social Science at City University of Hong Kong with a joint-appointment between Department of Media & Communication and School of Data Science. His current research focuses on the structure, content, use, and effects of social and mobile media. (ORCID: https://orcid.org/0000-0001-6173-6941)

**Mobile Phone Use as Sequential Processes:**

**From Discrete Behaviors to Sessions of Behaviors and Trajectories of Sessions**

**Abstract:** Mobile phone use is an unfolding process by nature. In this study, it is explicated as two sequential processes: mobile sessions composed of an uninterrupted set of behaviors and mobile trajectories composed of mobile sessions and mobile-off time. A dataset of a five-month behavioral logfile of mobile application use by approximately 2,500 users in Hong Kong is used. Mobile sessions are constructed and mined to uncover sequential characteristics and patterns in mobile phone use. Mobile trajectories are analyzed to examine intraindividual change and interindividual differences on mobile re-engagement as indicators of behavioral dynamics in mobile phone use. The study provides empirical support for and expands the boundaries of existing theories about combinatorial use of information and communication technologies (ICTs). Finally, the understanding on mobile temporality is enhanced, that is, mobile temporality is homogeneous across social sectors. Furthermore, mobile phones redefine, rather than blur, the boundary between private and public time.

**Mobile Phone Use as Sequential Processes:**

**From Discrete Behaviors to Sessions of Behaviors and Trajectories of Sessions**

The past decades have witnessed drastic and fundamental changes in the media ecosystem. Among these changes, the emergence of mobile phones is arguably one of the most disruptive ones. Mobile phones have been structured into people's everyday lives (Ling, 2012) due to the "always-on, always-on you" (Turkle, 2008) quality of mobile communication technologies. Mobile phones are perceived as an extension of its users; as such, the use of mobile phones has become personalized and customized (Campbell, 2013). Unraveling behavioral patterns on mobile phones has been a prominent topic in communication research and has inherited theoretical frameworks and methodological instruments from media use research of the past decades (Kim et al., 2017). Despite relevant insights into mobile phone use in previous studies, the increasing integration of mobile phones into people's lives and the increasing sophistication of mobile phone use require a new perspective to advance the understanding of behavioral patterns and behavioral dynamics in mobile phone use, and this perspective can improve theoretical understanding and empirical analysis of psychological and social consequences associated with mobile phone use.

**Discreteness and Multiplicity Frameworks in Media Use Research**

The two major theoretical frameworks in media use research are discreteness and multiplicity. The discreteness framework assumes that individuals make isolated choices about a discrete medium rather than a combination of media (Walther & Parks, 2002). Theoretical perspectives under such framework aim to understand why and how a particular medium will be chosen for use among a wide array of available media technologies (Stephens, 2007). The discreteness framework has heavily influenced empirical studies of mobile phone use. Empirical studies (e.g., Boase & Ling, 2013; Abeele, Beullens, & Roe, 2013; Wei, 2014) have used self-reported and logfile data to examine how mobile phone users engage with discrete behaviors (e.g., texting, calling, social networking, playing games) on mobile phones. With the proliferation of media options available to ordinary users, the discreteness framework has become insufficient to capture media use in realistic settings (Stephens, 2007). Ordinary users are enabled to select and use multiple media in a personalized way, thereby leading to the emergence of a multiplicity framework in media use research. The multiplicity framework aims to understand why and how multiple media are selected and combined for use.

Compared with the discreteness framework, the multiplicity framework provides a more valid approach to understand mobile phone use in real-life settings. Mobile phones as a multifunctional device have become a key entry point to the chromatic online world (Pew Research Center, 2015). Driven by progressively diverse personal needs and facilitated by increasingly convenient mobile technologies, mobile phone users have increased motivation and empowerment to engage with multiple behaviors on a single device. Mobile phone users can develop a self-defined repertoire of behaviors (Taneja et al., 2012) out of all available behaviors on mobile phones, ranging from computer-mediated communication to web surfing, social networking, playing games, and many others. Multiple behaviors in an individualized repertoire can be organized and combined by mobile phone users in two ways. The first is a concurrent combination of multiple behaviors known as multitasking, which has been empirically studied in recent years (e.g., David et al., 2015; Chen & Yan, 2016). The second is a sequential order of multiple behaviors, which is a common behavioral pattern in media (Leonardi et al., 2012) or ICT use (Stephens, 2007). Nevertheless, only few empirical studies have been conducted to understand mobile phone use as a sequential process.

The present study aims to fill in the research gap by explicating mobile phone use as two sequential processes, namely, mobile sessions and mobile trajectories. The sequential explication of mobile phone use enables researchers to understand human behaviors in continuity and capture the dynamics of mobile phone use. Particularly, the study aims to examine how individual users who are assumed to be rational and active agents select and organize discrete behaviors into mobile sessions and uncover sequential characteristics and patterns embedded in mobile sessions. Furthermore, the study explores users' behavioral dynamics underlying the alternation between mobile sessions and mobile-off time in mobile trajectories.

### From Discrete Behaviors to Sessions of Behavior and Trajectories of Sessions

Mobile phone use is an unfolding process over time. In such a sequential process, one type of behavior on mobile phones is contextualized from the network of other behaviors and prior times. An individual user transitions from one behavior to the next when he/she internally senses that the former behavior is complete, and the next behavior is relevant to and compatible with the former (Avnet & Sellier, 2011). The sequential explication of mobile phone use highlights users' rational decision making in the selection, order, and transition of behaviors on mobile phones. It gives rise to two sequential processes beyond discrete behaviors in mobile phone use, namely, sessions of behavior and trajectories of sessions.

### Sessions of Behaviors: Uncovering Sequential Characteristics and Patterns

The media repertoire theory (Watson-Manheim & Bélanger, 2007) argues that ordinary users can exert their locus of control to develop a personalized repertoire of discrete behaviors in media use. Moreover, each discrete behavior included in a self-defined repertoire is not independent and distinct from one another. To satisfy users' daily needs, discrete behaviors can be organized into a consecutive uninterrupted sequence of behaviors, which has been conceptualized as mobile sessions in empirical studies (Zhu et al., 2018).

Mobile sessions can possess different characteristics and demonstrate various patterns. Some sessions can include various consecutive behaviors with either complementary or redundant functions (Stephens et al., 2008; Bélanger & Watson-Manheim, 2006) to complete tasks with varying degrees of complexity in their work or lives. Meanwhile, other sessions can include one discrete behavior to which a user will stick in a certain timespan. Some behaviors embedded in a mobile session can endure, whereas others in the same session can be quite ephemeral. Uncovering sequential characteristics underlying mobile sessions is the first research question (RQ1) of this study.

Particularly, the following sequential characteristics, which are fundamentally important in the empirical understanding of sequential processes, are considered: duration, repertoire size, and order and transition. *Duration* is the most intuitive characteristic of mobile sessions and the most widely studied characteristic of mobile phone use in empirical research. In mobile sessions, duration refers to the time length of an overall session as well as the time length of specific behaviors in a session. *Repertoire size* refers to the number of discrete behaviors included in mobile sessions adopted by a user. Few users allocate their time to consume all available behaviors on mobile phones. Instead, most users develop a subset of behaviors out of available options and only engage with behaviors included in this subset (Watson-Manheim & Bélanger, 2007; Taneja et al., 2012). Users can differ in the size of their behavioral repertoires.

*Order and transition* among discrete behaviors in mobile sessions address questions on how mobile sessions initiate, develop, or terminate over time. Initiating behaviors in mobile sessions are important in communication research because they provide fundamental information about the driving forces underlying mobile phone use. Moreover, the transition between discrete behaviors in mobile sessions reflects the perceived compatibility or complementarity among multiple discrete behaviors, thereby enabling ordinary users to combine multiple behaviors into one complete session. According to previous studies on the combinatorial use of communicative media (Bélanger & Watson-Manheim, 2006) and ICTs (Stephens, 2007), two types of transitions can exist in mobile sessions: assortative and disassortative transitions. The former refers to a transition originating from and terminating on the same type of behavior in mobile sessions, whereas the latter refers to a transition originating from one type of behavior (e.g., texting) and terminating on another type of behavior (e.g., search) in mobile sessions. The present study aims to discover which initiating behavior is likely to occur in mobile phone use and which types of transition are popular in mobile sessions.

Furthermore, detecting any patterns underlying the selection, order, and transition of discrete behaviors in mobile sessions as a whole is theoretically desirable. The question regarding patterns focuses on global similarities to establish sequential typologies of mobile sessions for mobile phone users. In prior studies, the theoretical sequentiality of mobile phone use has been reconciled with the nonsequential measurements, such as the duration and frequency of mobile phone use. The consideration of the sequential order of human behaviors on mobile phones can provide an additional level of information on whatever behavior is being observed, a level that is inaccessible to a nonsequential perspective (Bakeman & Gottman, 1986). Therefore, the second research question (RQ2) in this study aims to uncover sequential patterns in mobile sessions initiated and generated by ordinary users of mobile phones.

**Mobile Trajectories: Understanding Behavioral Dynamics in Mobile Phone Use**

Individual mobile sessions can be regarded as paced, meaningful, and high-order sequences of events (Ancona et al., 2001a). It can be organized into a high-order sequential process in mobile phone use, namely, trajectories of mobile sessions (referred to as *mobile trajectories* hereinafter). Mobile trajectories amalgamate mobile sessions adopted by a user with their interposing mobile-off time into one sequential process. It offers a distinct perspective, that is, mobile re-engagement focusing on the alternation between mobile-off time and mobile sessions, to understand behavioral dynamics in mobile phone use.

Instead of being fully hooked with mobile phones in their daily lives, most ordinary users alternate between mobile-off time and mobile sessions. They withdraw themselves from mobile sessions and engage in other activities at a certain time period. Then, they withdraw from their mobile-off time and re-engage with their mobile phones. Mobile phone users are expected to develop various mobile re-engagement patterns at different time windows in a day to cope with their private and public affairs, and these patterns can fairly represent how individual users organize their private and public time (Zerubavel, 1979) in the new mobile age.

The presence of mobile-off time in mobile trajectories and the sequential nature of mobile trajectories provide a broad perspective in understanding mobile phone use. Such perspective goes beyond mobile phones *per se* and contextualize mobile phone use with other mobile-off behaviors. Not only can it convey important information on the behavioral dynamics of mobile phone use but also unveil important characteristics about mobile temporality. Mobile temporality is a new type of temporality in human society concerned with the conventions that govern the way by which time is organized and constructed during the use of mobile phones and with how mobile phone use is embedded into everyday lives (Green, 2002).

Mobile re-engagement as an indicator of behavioral dynamics of mobile phone use evolves over time and varies among users with different characteristics. In other words, *intraindividual change* and *interindividual differences* are available on mobile re-engagement. Existing empirical studies have focused on understanding the interindividual differences on mobile phone use; in these studies, the interindividual mean score on mobile phone use has been interpreted as a user's trait score, and the intraindividual variability around the mean over time has been either completely unmeasured or assumed to be random (Wang, Hamaker, & Bergeman, 2012). However, user behavior can spontaneously or reactively change over time even if all external stimuli are held constant. The intraindividual change of mobile phone use over time carries important information on how individuals differ from one another (Bem & Allen, 1974). This information must then be interpreted within a broad framework of interindividual and contextual differences (Ram & Gerstorf, 2009). The third research question of the study (RQ3) is to examine the intraindividual change and interindividual differences on the mobile re-engagement embedded in mobile trajectories.

## Research Methods

Data are collected from a representative panel of approximately 2,500 users in Hong Kong. The study is approved by the Human Subjects Ethics Committee at City University of Hong Kong. Users in the panel are recruited by a marketing research firm for local media organizations. The observation period starts from July 2016 to November 2016. At the beginning of the observation period, user demographic characteristics, namely, gender, age group, education level, and occupation, are collected via an online survey. An on-phone meter is installed on the Android and iOS mobile phones of the enrolled users. The users are informed about and agree to the tracking and subsequent analyses of mobile app uses under anonymity.

This meter passively tracks the use of mobile applications (referred to as *apps* hereinafter). The meters only track the use of the active apps on screen and filter the others running in the background. The meters record the start and end time of the use of each active app on mobile phones. However, the meters do not record if the use of an app is triggered by user-customized notifications on mobile phones or not. During the five-month observation period, all users in the panel generate approximately 11.5 million records of discrete app use.

All personally identifiable information about the users and the specific name of all mobile apps are removed to protect user privacy. The generic category of each app is available and provides the basis to examine sequential characteristics and patterns in mobile phone use. A total of 20 generic categories of mobile apps are included in the study: Communication, e-Commerce, Education, Email, Entertainment, Fashion, Finance, Games, Lifestyle, Music, News/Information (labeled *News* hereinafter), Photo, Search, Social Network Sites (referred to *SNS* hereinafter), Texting, Tools, Travel, Video, Web, and Miscellaneous (referred to as *Misc* hereinafter). Gender is measured as a dichotomous variable. The reported actual age is recoded into five age groups (i.e., 18-20, 21-30, 31-40, 41-50, and 51-64). Education level is measured as an ordinal variable with six categories, which are collapsed into three levels (i.e., low, medium, and high). Occupation is measured as a discrete variable with six categories: managers, professionals, clerks, workers, students, and unemployed.

An individualized approach is used for all the data processing and analysis of the study, including the construction of mobile sessions, detection of sequential patterns, and analysis of mobile re-engagement. Mobile sessions are constructed from the original records of discrete app use with the inter-app interval as a threshold. The inter-app interval (Barabasi, 2005) refers to the elapsed time between two consecutive app uses. An individualized threshold, which is the median score of a user's inter-app intervals, is adopted for each user. If the inter-app interval is smaller than the median inter-app interval of a user, then the two apps are grouped into a mobile session; otherwise, they are assigned to two distinct sessions. In total, 5.5 million mobile sessions are constructed out of the 11.5 million records of discrete app use. Among the 5.5 million mobile sessions, 3.7 million (67%) are solo-app sessions, which involve the one-time (56%) and repeated (11%) use of one app category in a session. Meanwhile, the remaining 1.8 million (33%) are multi-app sessions referring to a sequential array of at least two app categories in a session. The time spent on solo-app mobile sessions only accounts for 31% of the total mobile time. Meanwhile, the remaining 69% is allocated to multi-app mobile sessions.

Second, the multi-app mobile sessions of each user are aligned to detect sequential patterns underlying multi-app sessions. Particularly, the similarity between multi-app mobile sessions is estimated for each user on the basis of a pairwise sequence comparison technique called optimal matching analysis between sequences of spells (Studer & Ritschard, 2016). Then, $k$-medoid cluster analysis (Kaufman & Rousseeuw, 1987) is conducted on the similarity matrix to cluster the multi-app sessions adopted by a user into $k$ clusters. The clustering analysis yields at least two clusters of mobile sessions for each user. In each cluster of multi-app sessions, a medoid session, whose average dissimilarity from all other sessions in the cluster is minimal, is identified. The medoid session is considered representative of all other sessions in a cluster. The medoid sessions of each user are then pooled together, thereby leading to a collection of 2,203 medoid sessions. These 2,203 medoid sessions are used as a measure of sequential patterns underlying the 1.8 million multi-app mobile sessions of all users. Similarity matrix estimation is implemented with TraMineR package (Gabadinho et al., 2011) in R. The $k$-medoid cluster analysis is implemented with WeightedCluster package (Studer, 2017) in R.

Third, the solo-app mobile sessions and medoid multi-app mobile sessions adopted by a user in a specific timespan are ordered and concatenated with the interposing mobile-off time into a mobile trajectory. Mobile re-engagement is measured as the probability to switch at a given position from $s_i$ (i.e., mobile-off time) to $s_j$ (i.e., a mobile session) in a set of mobile trajectories in a certain timespan. The switch rate $p(s_j|s_i)$ between states $s_i$ and $s_j$ is calculated as

$$p(s_j|s_i) = \frac{\sum_{t=1}^{L-1} n_{t,t+1}(s_i, s_j)}{\sum_{t=1}^{L-1} n_t(s_i)},$$

where $n_t(s_i)$ refers to the number of mobile trajectories that do not end in $t$ with state $s_i$ at position $t$, and $n_{t,t+1}(s_i, s_j)$ is the number of mobile trajectories with state $s_i$ at position $t$ and state $s_j$ at position $t+1$. $L$ is the maximal observed length of all mobile trajectories in a timespan. The possible value of mobile re-engagement ranges from 0 to 1.

Linear mixed modeling is performed to unravel the intraindividual change and interindividual differences on mobile re-engagement. To examine the intraindividual change in mobile trajectories, the 24-hour cycle is divided into five timespans: *small hours* (0–07:59), *morning hours* (08:00–11:59), *midday hours* (12:00–13:59), *afternoon hours* (14:00–17:59), and *evening hours* (18:00–23:59). Small and evening hours are considered as private time and morning and afternoon hours are considered as public time. Midday hours are considered as middle ground between private time and public time. One type of intraindividual change, namely, circadian rhythm, and four interindividual differences, namely, gender, age group, education level, and occupation, are considered in the model. Linear mixed modeling is implemented with lme4 package (Bates et al., 2014) in R.

**Analytical Findings**

Analytical findings are organized as follows: First, sequential characteristics in mobile phone use are reported by analyzing 5.5 million constructed mobile sessions. Then, sequential patterns underlying multi-app mobile sessions are described by analyzing 2,203 medoid sessions. Finally, the intraindividual change and interindividual differences on mobile re-engagement are presented.

**Characterizing Mobile Sessions: Duration, Repertoire Size, Order and Transition**

On average, a user engages in 47 mobile sessions per day ($Mdn = 37$, $SD = 45$). The average duration of mobile sessions is 262 seconds ($Mdn = 74$ seconds, $SD = 562$). The session duration is positively correlated with the number of transitions in mobile sessions ($r = 0.54$, $p < 0.001$). As the use of each app category is now embedded in a mobile session, app-specific durations are found to be correlated with the number of transitions in mobile sessions. For most of the app categories, their app-specific duration in a mobile session is negatively correlated with the number of transitions in the session. However, in the music app, the durations within a session are positively correlated with the number of transitions in the session.

The average repertoire size in mobile phone use is 15 ($Mdn = 15$, $SD = 4$). In other words, individual users of mobile phones adopt 15 categories of mobile apps on average. Less than 1% of the users are omnivorous users who exhaust all 20 app categories in their behavioral repertoire. Meanwhile, approximately 10% are highly selective users who include less than 10 app categories in their repertoire.

Users prefer to launch a multi-app mobile session with Communication, Tools, SNS, and Search apps. Approximately 34.7% of the initiating apps in multi-app sessions are Communication, followed by Tools (15.7%), SNS (12.3%), and Search (6.9%). The second unique app after the initiating app in multi-app sessions is dominated by SNS (25.2%), Communication (18.8%), Search (13.7%), and Tools (9.3%). In solo-app sessions, the top five are Communication (16.3%), Tools (10.2%), SNS (9.4%), Games (5.6%), and Search (5.2%).

The average number of transitions in mobile sessions is 1 ($SD = 2$). In mobile sessions, assortative transitions substantially outweigh disassortative ones. On average, the top five app categories with the greatest assortative transition rates are Games, SNS, Search, Communication, and Video. The most frequently occurring disassortative transitions in mobile sessions are Email → Search, Texting → Communication, Texting → SNS, Photo → Communication, and Texting → Tools. In disassortative transitions, the top five app categories with the greatest outgoing rates are Search, Tools, Entertainment, SNS, and Communication. Meanwhile, the top five app categories with the greatest incoming rates are Tools, Communication, Search, SNS, and Games.

**Mapping Sequential Patterns underlying Multi-app Mobile Sessions**

Out of the 1.8 million multi-app mobile sessions generated by all users during the observation period, 2,203 sequential patterns are extracted. The distribution of the sequential patterns is highly skewed with the top 30 sequential patterns representing 50% of the 1.8 million multi-app sessions. As shown in Figure 1, the top 30 sequential patterns differ in their overall durations, component apps, and duration and order of each component app. Only seven categories of apps, namely, Communication, Email, Games, News, Search, SNS, and Tools, are included in the top 30 sequential patterns, except for Misc, although 20 categories of apps are available for users to select and combine.

[Figure 1 about here]

 "Communication and SNS" and "Communication and Tools" are the most popular app pairs combined by mobile phone users. As shown in Figure 1, among the top 10 sequential patterns, four patterns are an array of Communication and SNS apps, and another four patterns are an array of Communication and Tools apps. Search is included in seven popular sequential patterns in which a Search app follows the use of other app categories, such as Communication,

SNS, Tools, and Email. The Games app is incorporated in five popular sequential patterns. It is heavily used in combination with a light use of another app category. The News app is used heavily following a light use of Communication in the top 16th sequential pattern. Meanwhile, Email is used lightly, followed by a heavy use of Search in the top 22nd pattern.

Some differences on the popular sequential patterns can be observed across gender and age groups, as depicted in Figures 2 and 3 respectively. Among the top 10 popular sequential patterns, male users have two patterns terminating with a heavy use of the Games app, which originates from a short use of the Communication and Tools apps, respectively. Female users do not have any sequential patterns including the Games app among their top ten patterns. However, female users have two sequential patterns that originate from an SNS app and terminate with a Communication app. By contrast, male users do not have any sequential patterns starting with an SNS app among their top 10 sequential patterns.

[Figure 2 and Figure 3 about here]

Among the top 10 popular sequential patterns, young users (i.e., 18–20, 21–30, and 31–40) have a pattern with two transitions, that is, starting and ending with a light use of a Communication app, including a heavy use of SNS in the middle. Meanwhile, old users (i.e., 41–50 and 51–64) do not have such a pattern. Users in the age groups of 21–30 and 31–40 have one sequential pattern, that is, starting with a light use of a Communication app and ending with a heavy use of a Games app. Meanwhile, users in the other three age groups do not have any patterns with the Games app included among the top 10 popular patterns. Users in the age group of 41–50 have a sequential pattern starting with a light use of an Email app and ending with a heavy use of a Search app. Users in the oldest age group (i.e., 51–64) have two sequential patterns that do not occur in the top 10 patterns among users in the other four age groups: one is a sequential pattern originating with a light use of a Communication app and terminating with a heavy use of a News app and the other one starting and ending with a Video app, with use of an Entertainment app in between.

**Intraindividual Change and Interindividual Differences on Mobile Re-engagement**

The circadian rhythm of mobile re-engagement demonstrates an inverse U shape. Users have the lowest re-engagement rate during the small hours. This rate shows an upward trend in the morning hours and reaches the daily peak in the midday hours. Then, the mobile re-engagement rate displays a downward trend in the afternoon hours and further decreases in the evening hours.

Males do not differ significantly from females on their re-engagement rates. Nonetheless, the rate of mobile phone re-engagement among younger users is higher than that of the older users. Users in the two youngest age groups (18–20 and 21–30) have the greatest re-engagement rate, followed by those in the age groups of 31–40, 41–50, and 51–64. Users with medium education level have the greatest re-engagement rate. Meanwhile, users with high education level do not differ significantly from those with low education level in terms of their re-engagement rate. The effect of occupation on mobile re-engagement is reported in the Online Supplement.

[Figure 4 about here]

The circadian rhythm of mobile re-engagement rate significantly varies across gender, age groups, and education levels, as shown in Figure 4. Male users have greater mobile re-engagement rate compared with female users in the morning and evening hours. Nonetheless, male and female users do not differ significantly from one another in the small, midday, and afternoon hours. During small hours, users with low education level outperform users with medium and high education level in their mobile re-engagement rates. Nevertheless, in the other four timespans, users with medium education level have the greatest re-engagement rate, followed by users with high and low education level.

Users in the five age groups demonstrate a consistent inverse U shape in the circadian rhythms of mobile re-engagement rates. Younger users (i.e., 18–20 and 21–30) have significantly greater mobile re-engagement rates compared with older users (i.e., 41–50 and 51–64) in all five timespans. However, some significant and subtle differences across age groups can be observed. First, the difference on mobile re-engagement among the users of the five age groups during small hours is the smallest and becomes greater with the unfolding of time in a day. The difference reaches the maximum during evening hours when users in all five age groups differ significantly from one another in terms of their re-engagement rate. During the evening hours, the youngest users have the greatest re-engagement rate, whereas the oldest users have the lowest. Second, middle-aged users (i.e., 31–40) demonstrate a substantial change on their mobile re-engagement rates across five timespans of a day. In small hours, middle-aged users have a significantly lower re-engagement rate compared with the two younger groups but greater re-engagement rate compared with the two older groups. In the morning and midday hours, the re-engagement rate of middle-aged users increases to a level that is not significantly different than that of the two younger groups. In the afternoon hours, the re-engagement rate of this middle-aged group becomes significantly lower than that of the two younger groups. The gap on mobile re-engagement rate between middle-aged and younger users is further widened in evening hours.

### Conclusions and Discussion

This study is among the first studies that have attempted to examine mobile phone use as sequential processes among a representative panel of ordinary users in a society. The sequential explication of mobile phone use, time-stamped behavioral data, and multilevel analytical design used in the study result in some expected and interesting findings. These findings make theoretical and methodological contributions to the literature of media use research and sociology of time research.

First, the study offers fresh and rich insights into the intricate and multiplex behavioral patterns initiated and developed by users in their daily mobile phone use; this study also provides empirical support for and expands the boundaries of existing theories on the combinatorial use of ICTs. Second, this study proposes the sequential explication of mobile phone use that recognizes the inescapability of time and timing in human behavior (Langley et al., 2013). In this manner, this study enables the empirical examination of how mobile phone use is embedded in the daily lives of users, which improves empirical knowledge about mobile temporality substantially. Finally, the study demonstrates the potentials of the sequential perspective as a sharpened lens in observing media use and human communication processes.

**Beyond Discrete Behaviors: Enriched Understanding of Mobile Phone Use**

The combinatorial use and sequential pairing of ICTs have attracted increasing scholarly attention in the past decade. This study pushes this burgeoning line of research past its current focus on organizational and interpersonal communication toward a wide social context. With behavioral data collected among a representative sample in Hong Kong, the findings of the study provide empirical support to the existence of the combinatorial use of multiple media options in real-life contexts. The study broadens the understanding of how mobile apps can be and are used by ordinary users in a further realistic way (Stephens, 2007). The sequential explication of mobile phone use, which has been rarely explored in prior studies of sequential use of ICTs, drives the understanding of multiplex behavioral patterns on mobile phones into a conceptual terrain of activity, temporal ordering, fluidity, and change (Langley et al., 2013). The initiating apps identified in the study underline that social (i.e., Communication and SNS apps), utilitarian (i.e., Tools app), information (e.g., Search app), and entertainment needs (e.g., Games app) are the major driving forces to launch a mobile session.

The assortative and disassortative transitions between app categories in mobile sessions imply that an enriched understanding of perceived media characteristics should be developed in media use theories, such as the media richness (Daft & Lengel, 1986) and media affordance (Gibson, 1986) theories. Instead of perceiving each discrete medium separately, the findings of this study suggest that the traits or affordances associated with each discrete medium could be remixed or combined in an intended or unintended way by ordinary users to combine discrete media for satisfying their multiplex needs. User-initiated remixture or combination can lead to an exponentially expanding pool of possible traits or affordances associated with a limited number of mobile apps or media options. The top 30 sequential patterns in multi-app mobile sessions built upon a combination of seven app categories (except Misc app category) verify this expansion at the behavioral level.

When users need to enhance a certain type of experience (e.g., pleasure-seeking) in mobile phone use, they will combine the use of apps of the same category (e.g., Games or Video) into a sequence, thereby producing assortative transitions in mobile phone use. When users are faced with a complicated scenario requiring an assembly of multiple apps with various traits or affordances, users will bundle apps of different categories (e.g., Email → Search, Photo → Communication) into a sequence, possibly leading to disassortative transitions in mobile phone use. Future studies can be conducted to go beyond the discreteness framework and investigate empirically how users combine their perceptions about discrete media and how such combinations at a perceptual level will lead to a combinatorial use at a behavioral level.

Furthermore, this study advances our knowledge about behavioral patterns underlying the combinatorial use of ICTs with a whole-sequence approach. The whole-sequence approach considers each session of sequential use of mobile apps as a whole, rather than as stochastically generated from point to point (Abbott, 1995). It preserves the process nature of mobile phone use well and considers the full spectrum of information incorporated in mobile sessions, such as the total and consecutive time spent on each behavior and the order and transition between different behaviors (Studer & Ritschard, 2016). Both regularity and variations are detected in the sequential use of mobile phones, as evidenced by the skewed distribution of extracted sequential patterns. A small number of patterns are blockbuster sequences, which are popular among most users. Most sequential patterns are niche sequences adopted by a limited number of users. The variation in the sequential use of mobile phones is validated by the observed differences on the sequential patterns between male and female users and among users in different age groups.

Gender difference on sequential patterns of mobile phone use can be explained by the difference in the characteristics between male and female users. On the one hand, male users are more task-oriented and competitive than female users; on the other hand, female users are more socially oriented and assistive than male users (Eagly & Karau, 1991). Therefore, male users are more inclined to include a heavy use of Games apps in a mobile session compared with female users; meanwhile, female users are more inclined to launch a mobile session with SNS apps compared with male users. Users in the three youngest groups were born after 1980 and had grown up surrounded by technology; as such, they are more comfortable using technology compared with users in the two older groups. Prensky (2001) labelled these younger users as digital natives and older users as digital immigrants. Digital natives are technology savvy and depend on technology as an essential component of their lives. Hence, they are more inclined to embrace Communication and SNS apps into sequential use of mobile phones. Digital immigrants carry over their behavioral legacy in other media platforms to their mobile phone use. Therefore, they are more inclined to incorporate News, Search, and Email apps in their mobile sessions compared with digital natives. The oldest user group (i.e., 51–64) among digital immigrants have more leisure time than users in the other four groups. Such leisure time drives them to develop a sequential pattern that combines Video and Entertainment apps.

**Homogenized Mobile Temporality and Redefined Temporal Boundaries**

By uncovering the intraindividual change and interindividual differences on mobile re-engagement, the study reveals how mobile phone use is built into everyday lives to contribute to the understanding of mobile temporality in two aspects.

*Homogenization of Mobile Temporality across Social Sectors*

The inverse U-shaped circadian rhythm in mobile re-engagement is robust across individuals with different characteristics, despite some subtle interindividual differences on intraindividual change. The robustness in the circadian rhythms of mobile re-engagement indicates that users across social sectors are quite homogenous in mobile temporality. The homogenization of temporality, which has been observed in empirical studies, is important for the function of a modern, industrialized, and rationalized society (Lewis & Weigert, 1981). It is also a major manifestation of the increasing bureaucratization of social accessibility and professional commitments in modern societies (Zerubavel, 1979). Our study is among the first studies to reveal such homogenization of mobile temporality empirically with behavioral data.

Detecting the homogenization of mobile temporality in Hong Kong is expected due to the high penetration rate of mobile phones. Hong Kong is one of the richest cities in the world with well-developed telecommunication infrastructures. Mobile cellular and the Internet have been widely adopted in Hong Kong, whose penetration rates were 250% for mobile cellular in 2017 and 87% for Internet in 2018 (ITU, 2018). Approximately 98% of all Internet users in Hong Kong go online via mobile devices (ITU, 2018). Mobile phones have been well integrated into most aspects of daily lives in Hong Kong. Therefore, ordinary users at various social sectors in Hong Kong develop homogeneous mobile temporality. Nevertheless, the study cannot uncover how the homogenization of mobile temporality begins to take shape due to the relatively short observation period. Mobile phones and other ICTs continue to reshape temporal experience and collective time consciousness (Nowotny, 1992), either keeping the homogenization of mobile temporality stable for a long period of time or introducing some turbulence to mobile temporality for certain social sectors. Future studies are needed to examine if such homogenization is reproducible in other societies and to understand what individual characteristics and structural factors drive the formation, persistence, and change of such homogenization.

*Temporal Boundaries between Private and Public Time: Blurred or Redefined?*

Time is "indispensable to the regulation of the social accessibility of modern individuals as well as to the maintenance of the partiality of each of their various social involvements" (Zerubavel, 1979, p. 40). By drawing a boundary between public and private time, individuals can manage their social accessibility and withdraw periodically from their public selves in public time into their private ones in private time. Nevertheless, the "anywhere, anytime" connectivity promised by mobile technologies (Green, 2002) can either blur or break the existing boundary between public and private time and disempower the management of the social accessibility of users (Prasopoulou et al., 2006).

The findings of this study imply that the boundary between private and public time in mobile temporality is conspicuous and redefined rather than blurred. First, the boundary between public and private time in mobile temporality is clear, as suggested by a noticeable and significant difference on the mobile re-engagement between private (i.e., small and evening hours) and public time (i.e., morning and afternoon hours). Low mobile re-engagement rates during private time indicate that temporal fragmentation caused by mobile phone use is less obvious during private time compared with that during public time. This result is consistent with the classical argument that "users have far more control over their accessibility to other social activities or social agents within private time than outside of it" (Zerubavel, 1979, p. 42).

Moreover, mobile phone users interweave mobile phone use into their existing temporal texture and redefine the boundaries between public and private time by adopting appropriate behavioral patterns of mobile phone use at certain time windows. The redefined temporal boundary is evidenced by the high mobile re-engagement rate during midday hours. To most ordinary users, midday hours are their "banana time" (Roy, 1959) when they can break up the monotony of a day and can temporarily suspend their association with their occupational and other social roles (Zerubavel, 1979). Midday hours can be considered a temporal middle-ground lying in between private and public time. Ordinary users fragment their midday hours into small pieces by alternating frequently between mobile sessions and mobile-off time. The fragmented midday hours suggest that ordinary users consider mobile phones as a punctuation device (Ancona et al., 2001b) to divert from their daily routines. In addition to the difference on mobile re-engagement in private and public time, future studies can compare behavioral complexity in mobile phone use during private and public time to further prove the emergence of a redefined boundary in mobile temporality.

**Methodological Contribution—Sequential Process as a Sharpened Temporal Lens**

Mobile phone use occurs in time and consumes time (Drucker, 1967). Communication researchers have been inclined to understand mobile phone use through a temporal lens by explicitly or implicitly incorporating time into the measurements of mobile phone use in empirical studies. However, time is accepted as commonly understood or as given in communication research (Chaffee, 1991). The taken-for-grantedness of the time concept has permeated into mobile communication research, thereby leading to the dominance of the time-budget approach in measuring mobile phone use. The time-budget approach reduces mobile phone use to a duration or frequency. The general or specific use of mobile phones is examined using this approach. This approach "excludes all other aspects of time that might simultaneously have a bearing on people's lives, and that people relate to, at any one moment" (Adam, 1990, p. 95) and underestimates users' locus of control in their mobile phone use.

The sequential explication of mobile phone use offers a sharpened temporal lens to capture a multidimensional meaning of time (Adam, 1990; Flaherty, 2003). It offers a rich set of temporal dimensions, such as timing, pace, rhythms, and transition (Ancona et al., 2001b), to characterize individuals' mobile phone use. Moreover, mobile phone users are not only *consumers* but also *owners* of their time. They are entitled to bend mobile phones to their own use (Biocca, 1988). Ordinary users will not only budget their time for various behaviors on mobile phones but actively construct their time (Sorokin & Merton, 1937) by generating an individualized repertoire of behaviors and arraying selected behaviors into different sequential processes. The sequential explication of mobile phone use enables communication researchers to recognize users' locus of control in mobile phone use and unveil behavioral patterns based on their socially constructed time. Moreover, such explication, which has been drastically altered by fast-developing mobile technologies, empowers communication researchers to develop an updated understanding of the way individuals live, work, communicate, and organize their activities (Orlikowski & Barley, 2001; Stephens, 2007).

**Limitations and Opportunities for Future Research**

The study has limitations that need to be addressed in future studies. By assuming that mobile phone use is a rational choice guided by controlled psychological processing, the study falls short of addressing the habitual use of mobile phones (Oulasvirta et al., 2011). Future studies with a long observation period can examine which sequential patterns are likely to become a habit. Secondly, the study disregards mobile phone use triggered by user-customized notifications on mobile phones. Future studies can examine how users perceive notification settings on mobile phones and if behavioral patterns triggered by notifications are different from those not triggered by notifications. Thirdly, the study does not consider the concurrent engagement of multiple behaviors. Future studies can examine the concurrent array of multiple behaviors in conjunction with the sequential array of multiple behaviors to provide a comprehensive examination of behavioral patterns in mobile phone use. Although the use of inter-app interval as a threshold to identify mobile sessions has its theoretical and empirical support, cross-validating the approach with alternative approaches, such as the one based on the on–off of screen, is desirable in future research. Moreover, the robustness of the findings in the study is tied to the categorization of individual apps. A certain proportion of apps are categorized into the Miscellaneous category based on the Google Mobile App Categories, which can be further mined with refined app categorization scheme. Future studies with different app categorization schemes are needed to assess the replicability of our findings.

**References**

Abbott, A. (1995). Sequence analysis: New methods for old ideas. *Annual Review of Sociology, 21*(1), 93-113. doi:10.1146/annurev.so.21.080195.000521

Adam, B. (1990). *Time and social theory*. Philadelphia: Temple University Press.

Ancona, D. G., Goodman, P. S., Lawrence, B. S., & Tushman, M. L. (2001a). Time: A new research lens. *The Academy of Management Review, 26*(4), 645-663. doi:10.2307/3560246

Ancona, D. G., Okhuysen, G. A., & Perlow, L. A. (2001b). Taking time to integrate temporal research. *The Academy of Management Review, 26*(4), 512-529. doi:10.2307/3560239

Avnet, T., & Sellier, A. L. (2011). Clock time vs. Event time: Temporal culture or self-regulation? *Journal of Experimental Social Psychology, 47*(3), 665-667. doi:10.1016/j.jesp.2011.01.006

Bakeman, R., & Gottman, J. M. (1986). *Observing interaction: An introduction to sequential analysis*. Cambridge, MA: Cambridge University Press.

Barabasi, A. L. (2005). The origin of bursts and heavy tails in human dynamics. *Nature, 435*(7039), 207-211. doi:10.1038/nature03459

Bates, D., Mächler, M., Bolker, B., & Walker, S. (2014). Fitting linear mixed-effects models using lme4. *arXiv preprint* arXiv:1406.5823.

Bélanger, F., & Watson-Manheim, M. B. (2006). Virtual teams and multiple media: Structuring media use to attain strategic goals. *Group Decision and Negotiation, 15*(4), 299-321. doi:10.1007/s10726-006-9044-8

Bem, D. J., & Allen, A. (1974). On predicting some of the people some of the time: The search for cross-situational consistencies in behavior. *Psychological Review, 81*(6), 506-520. doi:10.1037/h0037130

Biocca, F. A. (1988). Opposing conceptions of the audience: The active and passive hemispheres of mass communication theory In J. A. Anderson (Ed.), *Communication yearbook* (Vol. 11, pp. 51-80). Newbury Park: Sage.

Boase, J., & Ling, R. (2013). Measuring mobile phone use: Self-report versus log data. *Journal of Computer-Mediated Communication, 18*(4), 508-519. doi:10.1111/jcc4.12021

Campbell, S. W. (2013). Mobile media and communication: A new field, or just a new journal? *Mobile Media & Communication, 1*(1), 8-13. doi:10.1177/2050157912459495

Chaffee, S. (1991). *Communication concepts 1: Explication*. Newbury Park, CA: Sage.

Chen, Q., & Yan, Z. (2016). Does multitasking with mobile phones affect learning? A review. *Computers in Human Behavior, 54*, 34-42. doi:10.1016/j.chb.2015.07.047

Daft, R. L., & Lengel, R. H. (1986). Organizational information requirements, media richness and structural design. *Management Science, 32*(5), 554-571. doi:10.1287/mnsc.32.5.554

David, P., Kim, J. H., Brickman, J. S., Ran, W., & Curtis, C. M. (2015). Mobile phone distraction while studying. *New Media & Society, 17*(10), 1661-1679. doi:10.1177/1461444814531692

Drucker, P. F. (1967). *The effective executive*. Oxford, UK: Butterworth-Heinemann.

Eagly, A. H., & Karau, S. J. (1991). Gender and the emergence of leaders: A meta-analysis. *Journal of personality and social psychology, 60*(5), 685-710. doi:10.1037/0022-3514.60.5.685

Flaherty, M. G. (2003). Time work: Customizing temporal experience. *Social Psychology Quarterly, 66*(1), 17-33. doi:10.2307/3090138

Gabadinho, A., Ritschard, G., Mueller, N. S., & Studer, M. (2011). Analyzing and visualizing state sequences in r with traminer. *Journal of statistical software, 40*(4), 1-37. doi:10.18637/jss.v040.i04

Gibson, J. J. (1986). The ecological approach to visual perception. Mahwah, NJ: Erlbaum.

Green, N. (2002). On the move: Technology, mobility, and the mediation of social time and space. *The information society, 18*(4), 281-292. doi:10.1080/01972240290075129

Haythornthwaite, C. (2005). Social networks and Internet connectivity effects. *Information, Communication & Society, 8*, 125-147. doi:10.1080/13691180500146185

ITU (2018, December 7). ITU releases 2018 global and regional ICT estimates. Retrieved from https://www.itu.int/en/mediacentre/Pages/2018-PR40.aspx

Kaufman, L., & Rousseeuw, P. J. (1987). Clustering by means of medoids. In Y. Dodge (Ed.), *Statistical data analysis based on the l1-norm and related methods* (pp. 405-416). North Holland: Elsevier.

Kim, Y., Kim, B., Kim, Y., & Wang, Y. (2017). Mobile communication research in communication journals from 1999 to 2014. *New Media & Society, 19*(10), 1668-1691. doi:10.1177/1461444817718162

Langley, A., Smallman, C., Tsoukas, H., & Van de Ven, A. H. (2013). Process studies of change in organization and management: Unveiling temporality, activity, and flow. *Academy of Management Journal, 56*(1), 1-13. doi:10.5465/amj.2013.4001

Leonardi, P. M., Neeley, T. B., & Gerber, E. M. (2012). How managers use multiple media: Discrepant events, power, and timing in redundant communication. *Organization Science, 23*(1), 98-117. doi:10.1287/orsc.1110.0638

Lewis, J. D., & Weigert, A. J. (1981). The structures and meanings of social time. *Social Forces, 60*(2), 432-462. doi:10.2307/2578444

Ling, R. (2012). *Taken for grantedness: The embedding of mobile communication into society*: Cambridge, Massachusetts: MIT Press.

Nowotny, H. (1992). Time and social theory: Towards a social theory of time. *Time & Society, 1*(3), 421-454. doi:10.1177/0961463X92001003006

Orlikowski, W. J., & Barley, S. R. (2001). Technology and institutions: What can research on information technology and research on organizations learn from each other? *MIS Quarterly, 25*(2), 145-165. doi:10.2307/3250927

Oulasvirta, A., Rattenbury, T., Ma, L., & Raita, E. (2012). Habits make smartphone use more pervasive. *Personal and Ubiquitous Computing, 16*(1), 105-114. doi:10.1007/s00779-011-0412-2

Pew Research Center. (2015, April 1). The smartphone difference. Retrieved from http://www.pewinternet.org/2015/04/01/us-smartphone-use-in-2015/

Prasopoulou, E., Pouloudi, A., & Panteli, N. (2006). Enacting new temporal boundaries: the role of mobile phones. *European Journal of Information Systems, 15*(3), 277-284. doi:10.1057/palgrave.ejis.3000617

Prensky, M. (2001). Digital natives, digital immigrants part 1. *On the horizon, 9*(5), 1-6. doi:10.1108/10748120110424816

Ram, N., & Gerstorf, D. (2009). Time-structured and net intraindividual variability: Tools for examining the development of dynamic characteristics and processes. *Psychology and Aging, 24*(4), 778-791. doi:10.1037/a0017915

Roy, D. F. (1959). Banana time job-satisfaction and informal interaction. *Human Organization, 18*(4), 158-168.

Sorokin, P. A., & Merton, R. K. (1937). Social time: A methodological and functional analysis. *American Journal of Sociology, 42*(5), 615-629.

Stephens, K. K. (2007). The successive use of information and communication technologies at work. *Communication Theory, 17*(4), 486-507. doi:10.1111/j.1468-2885.2007.00308.x

Stephens, K. K., Sørnes, J. O., Rice, R. E., Browning, L. D., & Sætre, A. S. (2008). Discrete, sequential, and follow-up use of information and communication technology by experienced ICT users. *Management Communication Quarterly, 22*(2), 197-231. doi:10.1177/0893318908323149

Studer, M., & Ritschard, G. (2016). What matters in differences between life trajectories: A comparative review of sequence dissimilarity measures. *Journal of the Royal Statistical Society: Series A (Statistics in Society), 179*(2), 481-511. doi:10.1111/rssa.12125

Studer, M. (2017). Package 'weightedcluster'. Retrieved from https://cran.r-project.org/web/packages/WeightedCluster/WeightedCluster.pdf

Taneja, H., Webster, J. G., Malthouse, E. C., & Ksiazek, T. B. (2012). Media consumption across platforms: Identifying user-defined repertoires. *New Media & Society, 14*(6), 951-968. doi:10.1177/1461444811436146

Turkle, S. (2008). Always-on/always-on-you: The tethered self. In J. Katz (Ed.), *Handbook of mobile communication studies* (pp. 121-138). Cambridge, MA: MIT Press.

Walther, J. B., & Parks, M. R. (2002). Cues filtered out, cues filtered in: Computer-mediated communication and relationships. In M. L. Knapp & J. A. Daly (Eds.), *Handbook of interpersonal communication* (3rd ed., pp. 529–563). Thousand Oaks, CA: Sage.

Watson-Manheim, M. B., & Bélanger, F. (2007). Communication media repertoires: Dealing with the multiplicity of media choices. *MIS Quarterly, 31*(2), 267-293. doi: 10.2307/25148791

Wei, R. (2014). Texting, tweeting, and talking: Effects of smartphone use on engagement in civic discourse in China. *Mobile Media & Communication*, *2*(1), 3-19. doi: 10.1177/2050157913500668

Abeele, V. M., Beullens, K., & Roe, K. (2013). Measuring mobile phone use: Gender, age and real usage level in relation to the accuracy and validity of self-reported mobile phone use. *Mobile Media & Communication, 1*(2), 213-236. doi:10.1177/2050157913477095

Wang, L., Hamaker, E., & Bergeman, C. S. (2012). Investigating interindividual differences in short-term intraindividual variability. *Psychological Methods, 17*(4), 567-581. doi:10.1037/a0029317

Zerubavel, E. (1979). Private time and public time: The temporal structure of social accessibility and professional commitments. *Social Forces, 58*(1), 38-58. doi:10.2307/2577783

Zhu, J. J. H., Chen, H. X., Peng, T. Q., Liu, X. F., & Dai, H. X. (2018). How to measure sessions of mobile phone use? Quantification, evaluation, and applications. *Mobile Media & Communication, 6*(2), 215-232. doi:10.1177/2050157917748351

**Figure 1. Most Popular Sequential Patterns**

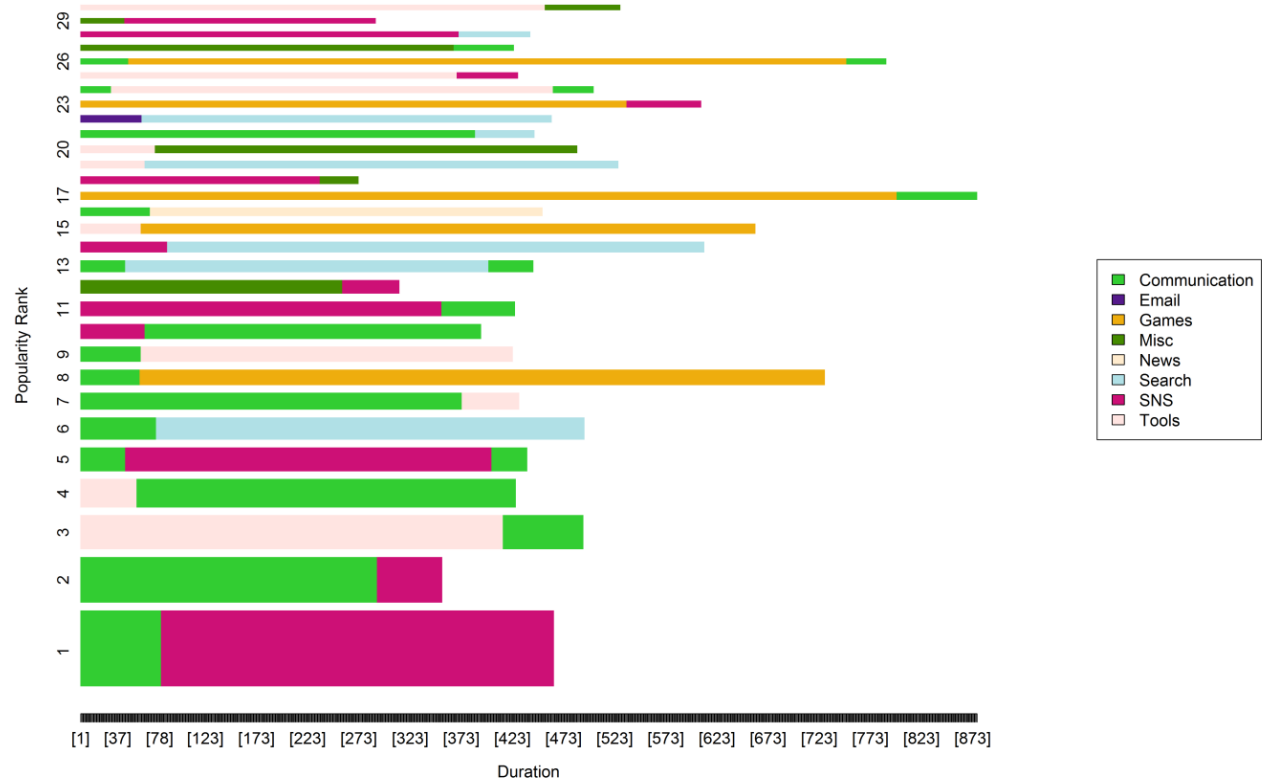

Notes: 1) The time unit of duration is in seconds. 2) The relative height of each pattern represents its popularity.

**Figure 2. Popular Sequential Patterns by Gender**

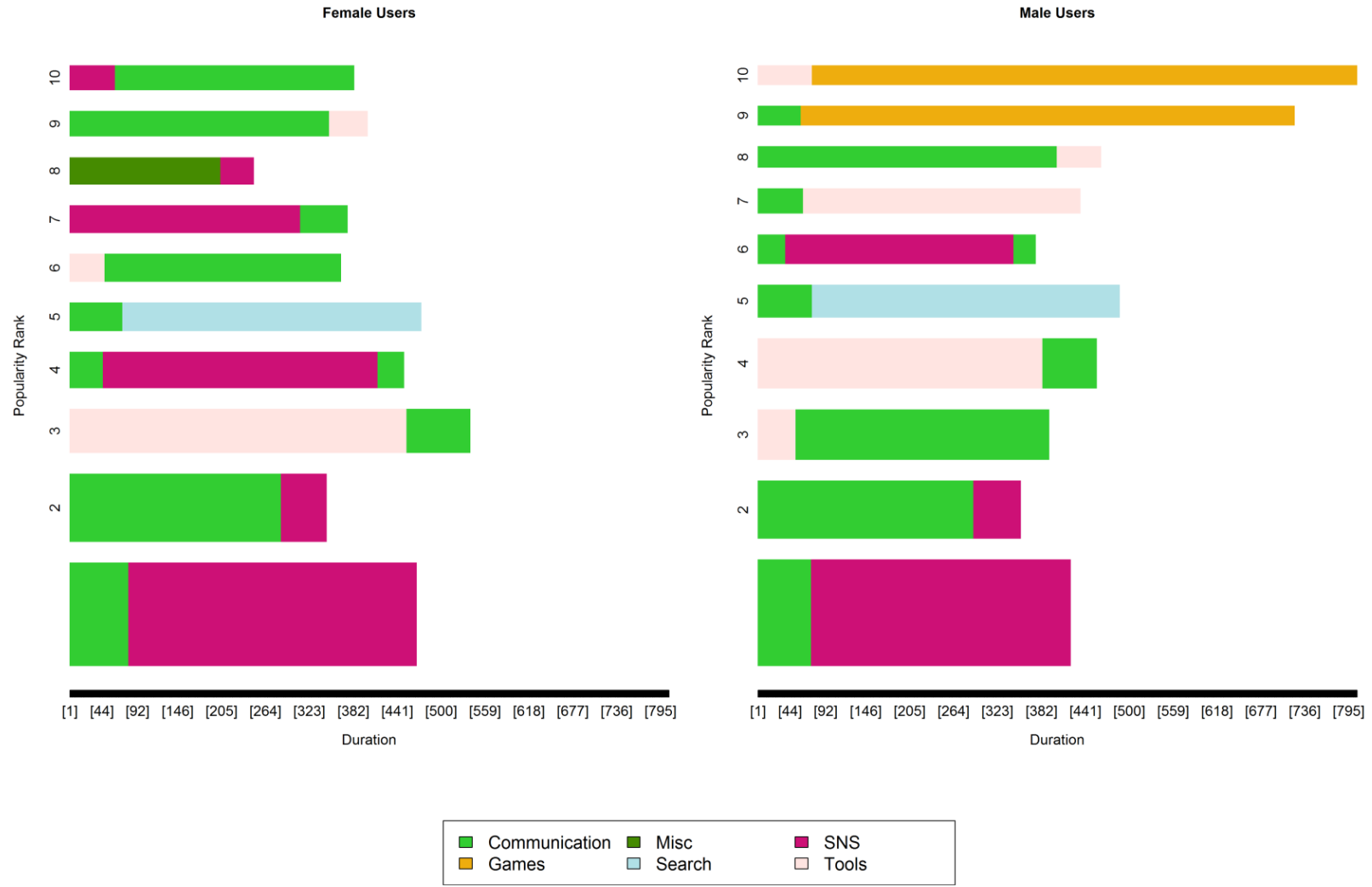

Notes: 1) The time unit of duration is in seconds. 2) The relative height of each pattern represents its popularity.

**Figure 3. Popular Sequential Patterns by Age Group**

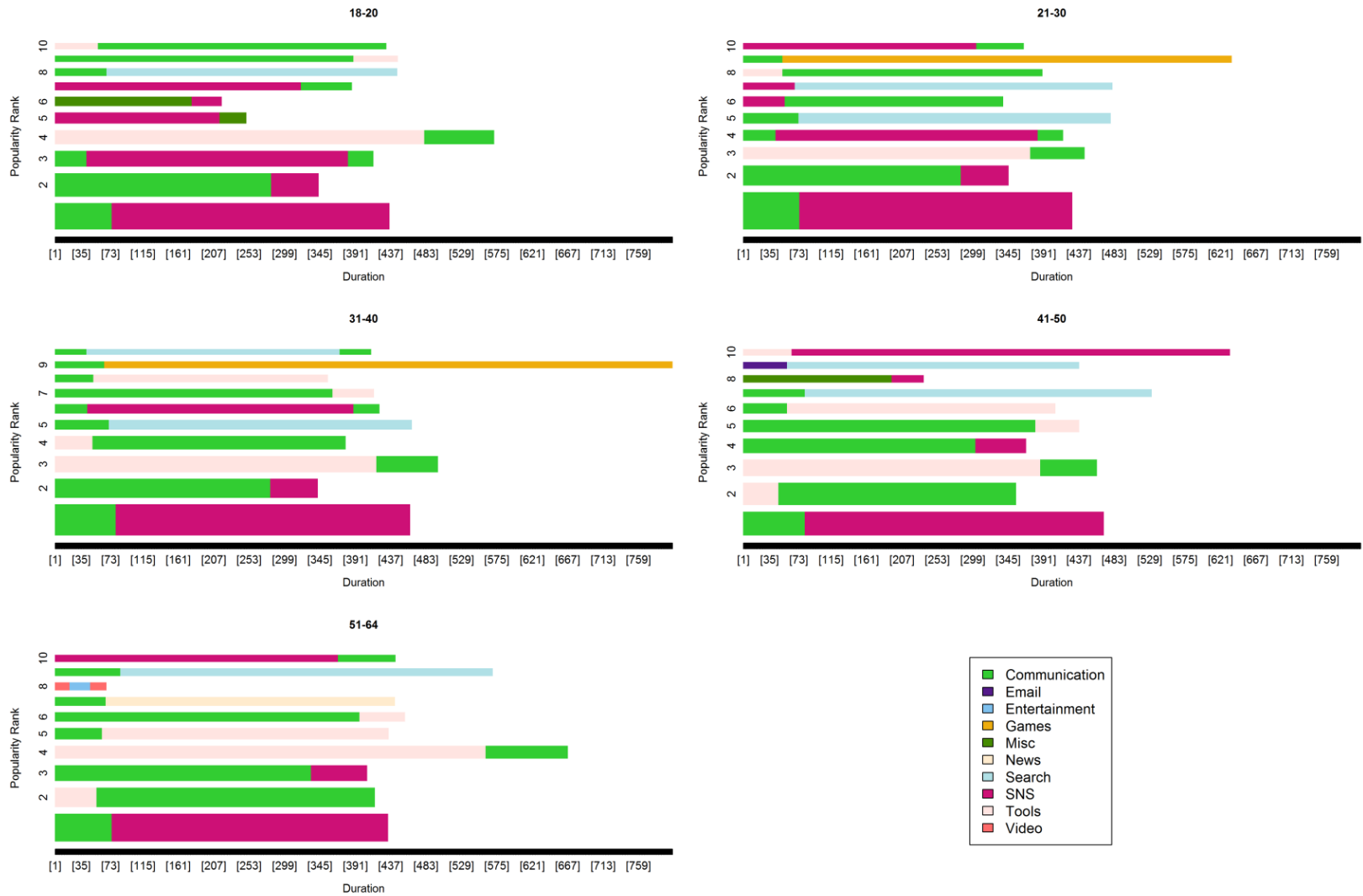

Notes: 1) The time unit of duration is in seconds. 2) The relative height of each pattern represents its popularity.

**Figure 4. Interindividual Differences on the Circadian Rhythm of Mobile Re-engagement**

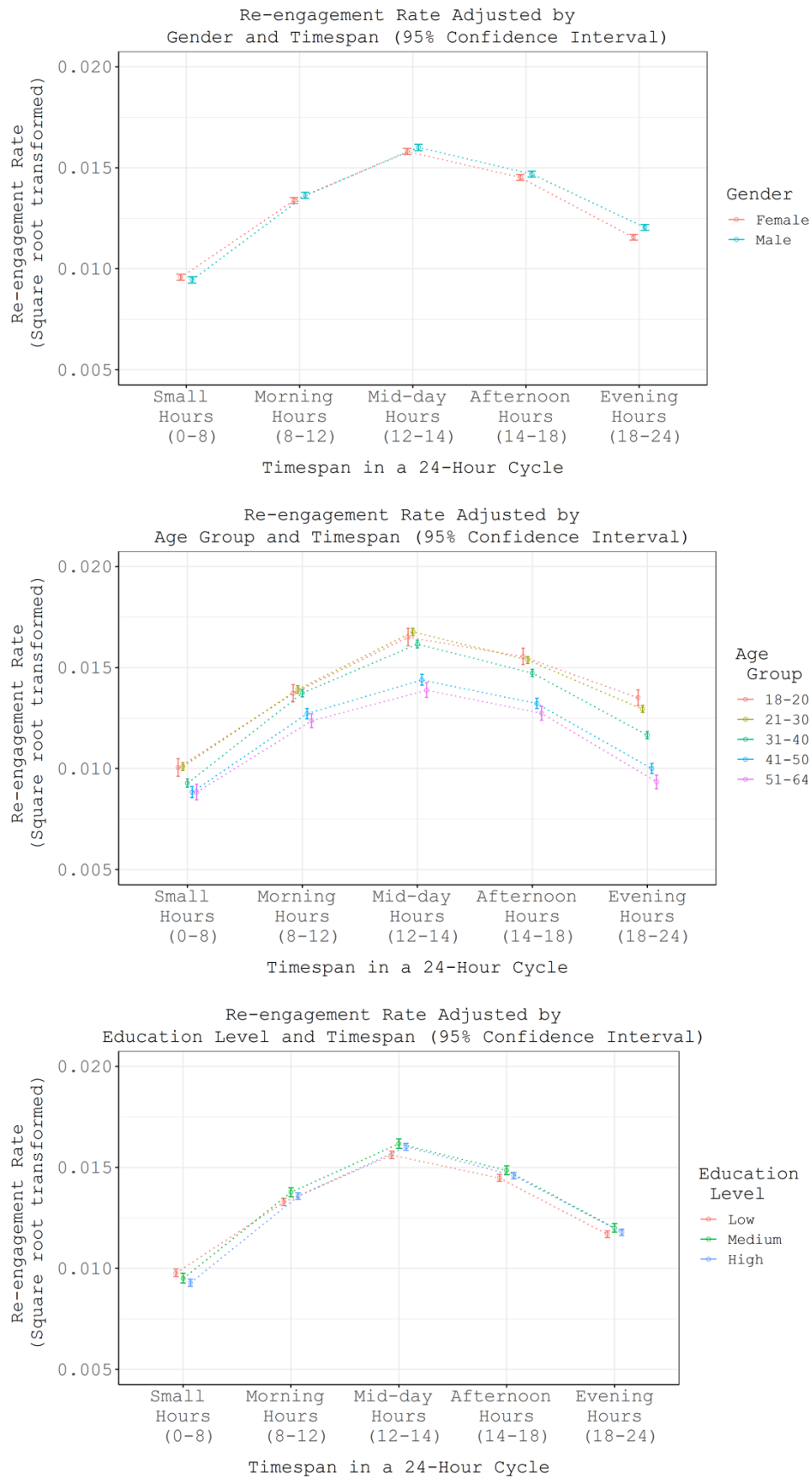